# The Transit Spectra of Earth and Jupiter

P.G.J. Irwin[a], J.K. Barstow[a], N.E. Bowles[a], L.N. Fletcher[a], S. Aigrain[b] and J.-M. Lee[c]

[a]Atmospheric, Oceanic, and Planetary Physics, Department of Physics, University of Oxford, Clarendon Laboratory, Parks Road, Oxford, OX1 3PU, United Kingdom.

*Tel:* (+44) 1865 272933, *Fax:* (+44) 1865 272923. E-mail: irwin@atm.ox.ac.uk

[b]Astrophysics, Department of Physics, University of Oxford, Clarendon Laboratory, Parks Road, Oxford, OX1 3PU, United Kingdom.

[c]Institute for Computational Science, University of Zurich, Winterthurerstrasse 190, CH-8057, Zurich, Switzerland.

Submitted to Icarus. Original: 12/5/14. Revised 24/7/14. Accepted 7/8/14.

Number of manuscript pages: 31 + acknowledgements, references, figure legends and tables (53 in total).

Number of Figures: 16

Number of Tables: 2

Editorial correspondence should be directed to:

Prof. Patrick G. J. Irwin,

Atmospheric, Oceanic and Planetary Physics, Clarendon Laboratory, Parks Road,

Oxford OX1 3PU, United Kingdom.

*Telephone:*   (+44) 1865 272933   (direct line: 272083)

*Fax:*   (+44) 1865 272923

*Email:*   irwin@atm.ox.ac.uk

**Abstract**

In recent years, an increasing number of observations have been made of the transits of 'Hot Jupiters', such as HD 189733b, about their parent stars from the visible through to mid-infrared wavelengths, which have been modelled to derive the likely atmospheric structure and composition of these planets. As measurement techniques improve, the measured transit spectra of 'Super-Earths' such as GJ 1214b are becoming better constrained, allowing model atmospheric states to be fitted for this class of planet also. While it is not yet possible to constrain the atmospheric states of small planets such as the Earth or cold planets like Jupiter, it is hoped that this might become practical in the coming decades and if so, it is of interest to determine what we might infer from such measurements. In this work we have constructed atmospheric models of the Solar System planets from 0.4 – 15.5 μm that are consistent with ground-based and satellite observations and from these calculate the primary transit and secondary eclipse spectra (with respect to the Sun and typical M-dwarfs) that would be observed by a 'remote observer', many light years away. From these spectra we test what current retrieval models might infer about their atmospheric states and compare these with the 'ground truths' in order to assess: a) the inherent uncertainties in transit spectra observations; b) the relative merits of primary transit and secondary eclipse spectra; and c) the advantages of acquiring directly imaged spectra of these planets. We find that observing secondary eclipses of the Solar System would not give sufficient information for determining atmospheric properties with 10m-diameter telescopes from a distance of 10 light years, but that primary transits give much better information. We find that a single transit of Jupiter in front of the Sun could potentially be used to determine temperature and stratospheric composition, but for the Earth the mean atmospheric composition could only be determined if it were orbiting a much smaller M-dwarf. For both Jupiter and Earth we note that direct imaging with sufficient nulling of the light from the parent star theoretically provides the best method of determining the atmospheric properties of such planets.

**1. Introduction**

The field of exoplanetary transit spectroscopy has advanced dramatically in recent years with the observed spectra of 'Hot Jupiter' planets such as HD 189733b and HD 209458b becoming increasingly better constrained. These spectra can be fitted with retrieval models to determine atmospheric states (Line et al., 2013; Lee, Fletcher and Irwin, 2012) and reveal atmospheres that are very different from anything seen in our Solar System. As the measurement techniques improve, the spectra of smaller, cooler 'Super- Earths' such as GJ 1214b (Barstow et al. 2013b; Benneke and Seager 2013; Kreidberg et al., 2014) are becoming measurable and ultimately planetary scientists will want to search the local galactic region for planets more similar to what we see in our Solar System and one day, perhaps, identify another Earth-like planet.

Should such a situation ever arise, it is of great interest to determine what we might actually deduce from the measured transit spectrum of a Solar System planet and a number of studies have been performed to investigate this. For example, Tinetti et al. (2006) modelled the disc-averaged spectrum of the Earth from $0.5 - 25$ μm, looking at the effect of various factors such as different surfaces and clouds, and more recently Rugheimer et al. (2013) have studied the disc-averaged spectra of Earth-like planets about F, G and K stars from 0.4 to 20 μm, looking at the visibility of detectable gaseous features. von Paris et al. (2013) looked to see how well potentially habitable planets could be characterized from secondary eclipse observations from proposed exoplanet missions such as the Exoplanet Characterisation Observatory (EChO, Tinetti et al. 2012), and also early direct imaging mission proposals such as Darwin (Léger et al., 1996; Cockell et al. 2009a; Cockell et al. 2009b). von Paris et al. (2013) found that using secondary eclipses or direct imaging the atmospheric composition would not be well determined, but that the surface temperatures of small rocky planets could be recovered reasonably well. For primary transits, Kaltenegger and Traub (2009) modelled the primary transit spectra of Earth-like planets about different stars from 0.3-20 μm looking at the detectability of different features and Bétrémieux and Kaltenegger (2013) examined the effect of atmospheric refraction and Rayleigh scattering in the UV to near-IR range.

In this work we construct simple atmospheric models of the Solar System planets Jupiter and Earth based on ground-based and satellite observations. From these models we compute their primary transit, secondary eclipse and directly-imaged spectra as seen from an observer ten light years away with a 10-m diameter space telescope, for the planets orbiting the Sun or an M-dwarf. We then examine what might be recoverable from these spectra by a 'remote observer' and compare the retrieved results with the actual atmospheric states of these planets.

**2. Construction and validation of Synthetic spectra**

For this study, synthetic spectra were calculated with the NEMESIS radiative transfer and retrieval model (Irwin et al. 2008). NEMESIS takes a model atmosphere, which defines the temperature profile as a function of pressure, together with the volume mixing ratio profiles of the constituent gases and the abundance profiles of clouds and aerosols and calculates the spectrum that one would expect to observe using, as a default, a correlated-k radiative transfer scheme. This modelled spectrum is then compared to that measured and the model parameters adjusted, using the technique of Optimal Estimation (Rodgers, 2000), to minimise the difference between the modelled and measured spectra. In this study we wish to simulate both thermal emission and reflected sunlight spectra of Earth and Jupiter and so we used a multiple scattering/thermal emission model, which is dealt with in NEMESIS using the matrix operator formalism of Plass et al. (1973); a five point Gauss-Lobatto quadrature scheme was chosen for the zenith angle integration, while the azimuthal integration was performed with Fourier decomposition, using N Fourier components, where N is set adaptively from the viewing zenith angle, $\theta$, as $N = int(\theta /3)$. To run this model, k-tables had first to be computed, which pretabulate the k-distributions of the absorption of different gases (e.g. Goody et al., 1989; Lacis and Oinas, 1991) at a set range of pressures and temperatures, for an assumed spectral resolution. In this study we chose to calculate these k-tables at a spectral resolution of 0.025 μm, to cover the range 0.4 to 15.5 μm. This resolution was chosen to be the same as the Galileo Near Infrared Mapping Spectrometer (NIMS) observations of Jupiter that we compare our model with (as described below) and is reasonably consistent with the spectral resolving power of the other observed spectra we used. The k-tables were calculated with 20 temperatures in the range 70 to 400K and 20 pressures equally spaced in log pressure between $3.1 \times 10^{-7}$ and 20.3 bar. Where available, the k-tables were calculated from HITRAN 2008 line data base (Rothman et al., 2008). However, for methane the band data of Karkoschka and Tomasko (2010) were used at near-IR wavelengths where the HITRAN 2008 data become insufficient. Similarly the $NH_3$ k-table was based on the band data of Bowles et al. (2008), combined with HITRAN 2004 and HITRAN 1996 linedata as described by Sromovsky and Fry (2010).

To construct the expected transit spectra of Earth and Jupiter, synthetic atmospheres first needed to be set up, with representative temperature, pressure, volume mixing ratio and cloud opacity profiles. These profiles were used to simulate the observations of Earth-observing and Jupiter-observing satellites to ensure consistency before moving on to simulate the transit spectra of these planets.

**2.1 Jupiter**

For Jupiter, the initial temperature/pressure/volume-mixing-ratio (vmr) profile was chosen to be consistent with the observations of the Composite Infrared Spectrometer (CIRS, Flasar et al., 2004) instrument on the NASA Cassini spacecraft, which covers the spectral range 7 – 1000 μm. A representative nadir co-added observed spectrum was used for this process and the atmospheric profiles fitted as described by Irwin et al. (2003) and Fletcher et al. (2009). To fit the cloud opacity needed to simulate the near-infrared and visible parts of the spectra, we made a simple approximation of assuming the particles to be composed of spherical droplets with a complex refractive index of 1.4 + 0i, and a standard Gamma size-distribution of mean-size 1 μm and variance 0.05. The extinction cross-section spectra and phase function spectra were then calculated with Mie Theory and the phase function approximated with combined Henyey-Greenstein functions for computational simplicity, where the phase function, $p(\theta)$, is modelled as

$$p(\theta) = \frac{1}{4\pi}\left[f\frac{1-g_1^2}{\left(1+g_1^2-2g_1\cos\theta\right)^{3/2}} + (1-f)\frac{1-g_2^2}{\left(1+g_2^2-2g_2\cos\theta\right)^{3/2}}\right]. \quad (1)$$

This function has three parameters, $f$, $g_1$ and $g_2$, where $g_1$ is the asymmetry of the forward scattering function, $g_2$ is the asymmetry of the backward scattering function and $f$ determines the relative contribution of each. Since the assumed particles have no absorption, the single-scattering albedo was calculated to be unity at all wavelengths. The scattering particles spectral properties were calculated with a step size of 0.1 μm, with linear interpolation between calculated points. Clouds and hazes in Jupiter's atmosphere were approximated with a single haze layer with variable base pressure and parameterised total nadir optical depth at 1.6 μm. The fractional scale height of this layer was set to 0.5, in accordance with NIMS near-infrared studies (Irwin et al., 2001). The synthetic model was compared with a set of four Galileo/NIMS spectra, previously analysed by Irwin et al. (1998), the so-called 'Real-time' spectra. Here we chose the fourth spectrum of this set, which has the highest 5-μm emission and thus the lowest opacity of the deeper cloud allowing radiation from the 5-8 bar level to escape to space, and adjusted the haze layer base pressure and optical depth to achieve reasonable agreement at near-IR wavelengths, eventually placing the haze at a base pressure of 0.56 bar, with a nadir optical depth at 1.6 μm of 5.25. The comparison between the synthetic spectra and the Galileo/NIMS and Cassini/CIRS spectra is shown in Figures 1 and 2, in terms of radiance and reflectivity respectively. Since the Galileo/NIMS and Cassini/CIRS spectra do not cover the complete spectral range of these simulations we also compared the calculations with the measured ISO/SWS (T. Encrenaz and T. Fouchet – private communication) spectrum (which is close to a disc-average) and with a reference visible ground-based albedo spectrum of Jupiter described by Karkoschka (1994). In the first panel of Figs 1 and 2, the synthetic spectrum is calculated at the Galileo/NIMS geometry of 42° solar incident angle and 0° emission angle (which is also consistent with the Cassini/CIRS observations), while the latter two panels are calculated at 45° solar incidence angle, 45° emission/reflected angle and 180° azimuth angle, i.e. in the back-scattering direction, which is more consistent with the geometry of the ISO/SWS and ground-based visible albedo observations (as discussed in section 3). It can be seen that good agreement is achieved between the modelled and measured spectra. One area of disagreement is in the 2.5-3.0 μm region, which is due to additional absorption by $NH_3$ ice and solid $NH_4SH$ (Irwin et al, 2001; Sromovsky and Fry, 2010). The poor performance at short visible wavelengths is due to the omission of the known blue/UV absorption of Jovian hazes. In both cases, the single-scattering albedo of our haze particles was adjusted by hand in these spectral regions to achieve the final improved fit shown in the third panels of Figs 1 and 2. Note that we tuned our model to give good agreement between the synthetic spectrum and the reference NIMS spectrum at 5 μm, but this agrees less well with the ISO/SWS spectrum. This is because the ISO/SWS spectrum is an average over cloudy and non-cloudy areas, whereas the reference NIMS spectrum was recorded in a region where the opacity of the deeper cloud was minimal.

**2.2 Earth**

For the Earth, the initial standard temperature/pressure profile was taken to be the US standard atmosphere of Anderson et al. (1986), which is typical for a latitude of 45.5°N. The surface temperature was set to the near-surface air temperature of 288.2K. A single cloud was chosen, with variable base altitude and variable opacity, with the fractional scale height set to 0.1. The complex refractive index spectra of the cloud particles was set to that of water vapour (Hale and Query, 1973), with again a standard Gamma size-distribution of mean-size 1 μm and variance 0.05 and the extinction cross-section, single scattering and phase function spectra (approximated with combined Henyey-Greenstein functions) calculated with Mie theory. To achieve a spectrum consistent with visible/near-IR observations, synthetic spectra were compared with Rosetta/VIRTIS (Coradini et al., 2007) observations of the Earth during Rosetta's third Earth flyby. The whole-disc observation of the Earth recorded by VIRTIS-M in the 00216741598 cube on 13$^{th}$ November 2009, was chosen for this exercise as it covers the entire disc and comprises of a representative sample of ocean, land and cloud. Since only half the disc was illuminated, just the lit side was included in the average spectrum. Clearly the geometry is rather different from the secondary eclipse case, but as a first approximation of the disc-averaged near-IR spectrum of the highly inhomegenous Earth, we judged this observation to provide a useful baseline. The resulting averaged spectrum was fitted with our retrieval model by simultaneously varying: the cloud opacity, the cloud base altitude, a coarse surface albedo spectrum (sampled only at every μm, with linear interpolation between), and a scaling factor of the assumed model water vapour abundance profile to attain a reasonable fit to the disc-averaged spectrum, which is shown in Fig. 3. For this fit, the cloud base height was retrieved to be at an altitude of 8.5 km, with an optical thickness (at 1.6 μm) of 0.45, while the retrieved surface albedo spectrum is shown in Fig. 4, compared with the albedo spectra of several different expected surfaces, such as the ocean, vegetation and desert, extracted from the JPL ASTER Spectral Library (http://speclib.jpl.nasa.gov/search-1). The fitted albedo spectrum in Fig. 4 appears to be a plausibly representative mean surface spectrum although it appears somewhat too bright at the very shortest wavelengths, suggesting some aliasing with the reflection from low-level clouds. It was not necessary to validate the mid-infrared spectrum as the typical temperature structure and composition of the Earth's atmosphere is very well known. This model was then frozen and used in all further calculations with the appropriately modified geometry.

When validating with the Rosetta/VIRTIS observation we found that we could achieve a good fit at near-IR wavelengths and near 5 μm (Fig. 3), but had less success in the 2.5-4.5 μm region, predicting more reflectance than is actually seen in this VIRTIS cube. We attribute this to the fact that the Earth is actually a very inhomogenous object, with variable clouds and variable surface properties, which is not well-approximated in this wavelength region by a single uniform cloud, single temperature profile, single water vapour profile and single albedo spectrum at all locations. For example, 'Earthshine' on the surface of the Moon reported by, among others, Montañés-Rodriguez et al. (2005), show the disc-averaged reflectivity of the Earth to be highly variable and Langford et al. (2009) found variations of up to 23% since the light is dominated by reflection from land, sea, cloud etc., whose relative contribution changes enormously as Earth rotates. However, the aim of this study is to assess the detectability of Earth and Jupiter's absorption and emission features in likely observed transit spectra rather than make a perfect simulation of the complicated spectrum of the Earth and all we need is a model atmosphere that generates a spectrum that is reasonably consistent with the observed spectra. Figure 3 shows that our model Earth atmosphere generates a spectrum that is sufficiently accurate for our purposes, but is also simple enough to easily test the retrieval accuracy in different observation scenarios.

## 3. Calculation of Synthetic Transit spectra

The NEMESIS model (Irwin et al. 2008) has recently been extended to model transit spectra and has been used for a number of recent analyses (Lee, Fletcher and Irwin, 2012; Barstow et al. 2013a,b, 2014). For primary transits, Nemesis calculates the limb transmission of the atmosphere, $T_R$, at different tangent altitudes and then integrates to find the total effective area of the planet at each wavelength:

$$A_P(\lambda) = \pi R_0^2 + \int_{R_0}^{\infty} 2\pi R \left(1 - T_R(\lambda)\right) dR \qquad (2)$$

where $R_0$ is the radius at a level well below the transparent part of the atmosphere, usually at the 10-bar level, or the solid surface for the Earth. The measured signal that is fitted by the model is then $100 \times A_P/A_S$, i.e. the percentage area ratio, where $A_S$ is the disc area of the host star. We should note that no account has been taken of the effects of atmospheric refraction in the current study. Studies such as Bétrémieux and Kaltenegger (2013) show that including refraction at visible wavelengths tends to slightly increase the continuum background absorption and thus reduce the visibility of gaseous molecular absorption features by two or three percent. Such effects are likely to diminish at longer wavelengths. Clouds are, of course, also important absorbers/scatterers and recently Robinson et al. (2014) have shown that neglecting refraction has minimal effect on modelling Titan transit spectra since the hazes limit the depths to which rays can pass through the atmosphere to altitudes not significantly affected by refraction. Since previous studies show the effect of omitting refraction has only a small effect on the Earth spectrum (over the entire range 0.4 – 16 μm considered here) and we know that Jupiter's atmosphere is hazy we believe that we are justified in not including refraction in this study. However, adding refraction could potentially be an avenue of future work.

For secondary eclipses in the case where reflected sunlight is negligible (e.g. at long wavelengths, or if the planet is very hot), NEMESIS assumes that radiative transfer can be approximated by a plane-parallel atmosphere and computes the emission into a hemisphere using exponential integrals (e.g. Goody and Yung, 1989). This is then integrated over the surface of the planet to give a total spectral power (W μm$^{-1}$) of:

$$P(\lambda) = 4\pi R_0^2 \times 2\pi \int_0^{0.5} B_\lambda(\tau_\lambda) dE_3(\tau_\lambda) \tag{3}$$

where $\tau_\lambda$ is the vertical optical depth at wavelength $\lambda$ and $E_n$ is the exponential integral:

$$E_n(x) = \int_1^\infty \frac{e^{-wx}}{w^n} dw. \tag{4}$$

NEMESIS then divides this power spectrum by the spectral power of the host star to give the power ratio $(P/P_{Star})_\lambda$, which is the same as the measured spectral flux (W m$^{-2}$ μm$^{-1}$) ratio $(F_P/F_S)_\lambda$.

For secondary eclipses in cases where reflected sunlight becomes significant, Eq.3 is not applicable and instead, the disc-averaged radiance (W m$^{-2}$ sr$^{-1}$ μm$^{-1}$) emerging from the atmosphere and travelling towards the observer must be integrated across the planetary disc:

$$\bar{R}_\lambda = \frac{1}{\pi R_0^2} \int_{r=0}^{R_0} \int_{\phi=0}^{2\pi} R_\lambda(r,\phi)\, r\, dr\, d\phi \tag{5}$$

where $r$ is the radius across the disc, $\phi$ is the polar angle of a position on the disc, $R_0$ is the effective radius of the planet, and $R(r,\phi)$ is the radiance emitted towards the observer from that position calculated with our multiple scattering model and thus including reflected sunlight. Expressing the radius across the disc in terms of the local zenith angle, $\theta$, Eq. 5 can be rewritten as:

$$\bar{R}_\lambda = \frac{1}{\pi} \int_{\theta=0}^{\pi/2} \int_{\phi=0}^{2\pi} R_\lambda(\theta,\phi) \sin\theta \cos\theta\, d\theta d\phi. \tag{6}$$

In reality the radiance emitted or scattered from the planet will vary with position. However, if we make the simplifying approximation of azimuth independence, then Eq.6 can be written:

$$\bar{R}_\lambda = \int_{\theta=0}^{\pi/2} R_\lambda(\theta) \sin 2\theta\, d\theta \tag{7}$$

which is the weighted mean of the radiance with the weighting function being sin2θ. Since the weighting function peaks at a zenith angle θ =45°, the disc-averaged radiance can be approximated further as $\bar{R}_\lambda \approx R_\lambda(45°)$. The spectral flux (W m$^{-2}$ μm$^{-1}$) arriving at the observer at a distance *D* from the planet is then:

$$F_\lambda = \frac{\pi R_0^2}{D^2} \bar{R}_\lambda \approx \frac{\pi R_0^2}{D^2} R_\lambda(45°) \tag{8}$$

which can be divided by the solar flux to give the flux ratio $(F_P/F_S)_\lambda$.

In this paper where we need to consider both reflected sunlight and thermal emission, Eq. 8 was used to calculate the planetary flux, assuming the same temperature/aerosol/abundance profile at all locations, using as mentioned earlier, a Matrix Operator multiple-scattering model (Plass et al., 1973), extended to include thermal emission. The radiance was calculated with both the solar and emission angles set to 45°, in the back-scattered direction – the geometry with which a planet is viewed immediately before and after a secondary eclipse. At longer wavelengths, where thermal emission dominates and scattering becomes insignificant, tests showed that model spectra calculated with this method and with the analytical hemispherical integration method (Eq.3) were effectively indistinguishable.

In this best case scenario study we assume that instrumental noise effects are minimal and thus that the measurements are photon noise limited. When measuring transit spectra we measure the total flux from the system before, during and after the transit and look for the small dip introduced by the planet on the overall spectral flux $F_\lambda$ (W m$^{-2}$ μm$^{-1}$). We calculate the noise spectrum using Eq.1 of Barstow et al. (2013a), to calculate the total number of photons detected during a given observation:

$$n_\lambda = F_\lambda \frac{\lambda}{hc} \Delta\lambda \, A_{eff} Q\eta t \tag{9}$$

where $\Delta\lambda$ is the full-width-half-maximum (FWHM) spectral resolution, $A_{eff}$ is the effective telescope area, $Q$ is the quantum efficiency of the detectors, $\eta$ is the throughput of the optical system (i.e. the overall transmission), and $t$ is the total integration time of the observation. If the system perfomance is noise-limited, then the noise on this incident photon flux is $\sqrt{n_\lambda}$ and the noise on the measured flux (W m$^{-2}$ µm$^{-1}$) is

$$\sigma_{F_\lambda} = \frac{F_\lambda}{\sqrt{n_\lambda}} \propto \sqrt{F_\lambda} \qquad (10)$$

since $n_\lambda \propto F_\lambda$. For secondary eclipses, the flux (W m$^{-2}$ µm$^{-1}$) measured during and before/after the transit is $F_1 = F_S$ and $F_2 = F_S + F_P$ respectively, where $F_S$ is the stellar flux and $F_P$ is the planetary flux. The planetary flux ratio is then extracted: $y = \frac{F_P}{F_S} = \left(\frac{F_2}{F_1} - 1\right)$. If $F_1$ and $F_2$ are measured with equal error $\sigma_{F_\lambda}$, then the error on y is $\sigma_y = \frac{\sqrt{2}\sigma_{F_\lambda}}{F_S} \propto 1/\sqrt{F_\lambda}$, and thus gets worse as the flux drops at longer wavelengths. For primary transits, the flux (W m$^{-2}$ µm$^{-1}$) measured before/after and during the transit is $F_1 = F_S$ and $F_2 = \frac{A_S - A_P}{A_S} F_S$ respectively. The signal extracted is the ratio of planetary area to the stellar area, i.e. $y = \frac{A_P}{A_S} = 1 - \frac{F_2}{F_1}$, and so the error of y is again $\sigma_y = \frac{\sqrt{2}\sigma_{F_\lambda}}{F_S}$, which once more gets worse at longer wavelengths. As we will see the $F_P/F_S$ signal for solar system planets is greatest at longer wavelengths, where the photon-limited noise performance is worse and so we might expect secondary eclipses to be less effective for determining the conditions in solar-system-like atmospheres. However, the $A_P/A_S$ signal for solar system planets shows features at all wavelengths, including those at shorter wavelengths where the photon-limited noise is reduced.

Thus we might expect primary transits to be more effective at probing solar-system-like atmospheres.

For this study we made the following assumptions. We assumed that the Solar System was being observed from a distance of 10 light-years, with a telescope of diameter 10 m. This distance was arbitrarily chosen to be far enough away to make direct-imaging difficult, but not so far away that the Sun becomes faint. The telescope diameter is typical of current ground-based telescopes and future planned, or proposed, space telescopes such as the James Webb Space Telescope (JWST) or the Advanced Technology Large-Aperture Space Telescope (ATLAST). We assumed that the throughput of the optical system is $\eta = 0.5$ and that the quantum efficiency of the detectors is $Q = 0.7$. These are typical figures for currently achievable infrared telescope systems and are the same as used in the EChO analysis of Barstow et al. (2013a). Furthermore, again following Barstow et al. (2013a), we assumed a duty cycle of 80%. As mentioned earlier, for our spectral resolution, we assumed a similar resolution of the Galileo/NIMS, Cassini/VIMS and Rosetta/VIRTIS instruments and chose a triangular instrument function with a FWHM = 0.025 μm.

## 4. Construction of a Retrieval Model

Assuming we had time to make transit observations with sufficient SNR, the second aim of this work was to establish how well the atmospheric states of these planets could be determined and compare these estimates to the known states of the Earth's and Jupiter's atmosphere used to generate the synthetic spectra.

The NEMESIS model was used to calculate the 'true' secondary eclipse flux ratio spectra of Earth and Jupiter, and Gaussian noise added to the level determined for a given number of transit observations. NEMESIS was then used to retrieve the atmospheric properties from a 'measured' spectrum and the retrieved atmosphere compared with 'true'. In our retrieval model, gases were in most cases assumed to be uniformly mixed (with *a priori* abundances equal to the abundances of each gas at the lowest level of the reference atmospheres, or representative stratospheric values for photochemically products such as ethane and acetylene) and the *a priori* error was set to 100%. Note that we retrieve log gas abundances in NEMESIS in order to prevent abundances ever becoming negative. In addition to the gases, a parameterised single cloud layer was included, with the fractional scale height fixed to 0.5, but the opacity and cloud base height allowed to vary. The *a priori* temperature profile was set to $T_{strat} = T_{eq}/2^{1/4}$ in the stratosphere (Irwin, 2009), where $T_{eq}$ is the radiative equilibrium temperature of the planet, down to an assumed tropopause at 0.1 bar (where the tropopause is typically found to occur for all the solar system planets, e.g. Robinson and Catling, 2014) and then followed a dry adiabatic lapse rate at higher pressures (assuming the molar heat capacity at constant volume, $C_v$=3R, consistent with an atmosphere dominated by polyatomic molecules). As we determine a continuous temperature profile, the *a priori* error needs to be tuned to allow sufficient freedom for the model to fit the temperature profile, but not too much that the solution becomes 'exact' and the model overfits the spectrum at the expense of allowing unrealistic oscillations to appear in the retrieved temperature profile. In addition, we know that in the gas giants, the temperature profile becomes barotropic at pressures greater than approximately 500 mbar, i.e. that the temperature/pressure profile becomes the same at all latitudes and longitudes due to the active convective overturning of such planets and thus this additional constraint was applied to our Jupiter a priori temperature profile. We assumed that for a giant planet such as Jupiter, the atmosphere would be expected to be dominantly composed of $H_2$-He and fixed this to near-solar values. For the Earth, we might initially suspect a $CO_2$ atmosphere, but the observed transit spectra (if measured with resonable precision) would quickly dispell this assumption and we would be led to the conclusion that the bulk constituent is transparent in the infrared, for which the most likely candidate would be a diatomic molecule such $H_2$, $O_2$, $N_2$, etc. $H_2$ could be discounted on exospheric escape grounds, while an atmosphere dominated by $O_2$ would be very unstably reactive. Hence, we would be drawn to consider a dominantly $N_2$-broadening atmosphere, which is what we have done here. It is possible that the molecular weight of an atmosphere (and thus bulk abundance) could be determined from the Rayleigh-scattering part of observed primary transit spectra, but only if we could be sure of cloud and haze-free conditions (e.g. Benneke and Seager, 2012; de Wit and Seager, 2013), which solar system experience would suggest is rather unlikely. Hence, we assumed the main consituent of Earth's atmosphere to be $N_2$ and looked at the retrievability of gases such as $H_2O$, $O_3$, $O_2$ etc.

We assumed that estimates of the planetary radius (from primary transits) are likely to be accurate to 5% since the star's radius is known to this precision from modelling its spectrum. Similarly, estimates of the planetary mass (from radial velocity motion of star) are also likely to be only accurate to 5% since the star's mass is only likely to be estimated to that accuracy. Hence, retrievals were performed not only with *a priori* profiles with the true mass and radius of Earth and Jupiter, but also with these perturbed by ±5% to quantify the effect of the uncertainty of these estimates on the retrieved atmospheric states. We should note that our assumed error on the planetary mass ignores the effects of stellar activity, would make it very difficult to detect the radial velocity modulations of an Earth-like planet around a Sun-like star. Hence, the initial error estimate of planetary mass might be actually be more like 20-30% for Earth-like planets. However, for the purposes of this simple study an error of 5% in planetary mass was tested for both Earth and Jupiter. Finally, for the primary transit cases we assumed two further cases where the stellar radius was ±5% different from the nominal assumption.

**5. Secondary Eclipse Simulations**

Using our model Earth and Jupiter atmospheres, we simulated the flux as seen from our observer at 10 light-years distance with a 10-m diameter telescope and compared these with the solar flux. Using the equations from the previous section we computed the photon-limited noise levels that could be achieved with 1, 1,000 and 1,000,000 hours of transit observations. Our comparisons are shown in Fig. 5. As can be seen, the photon noise on the solar flux completely swamps the signal of the planetary flux for any practically achievable integration time. At our original spectral resolution of 0.025 μm, we calculate that the required in-transit integration time to achieve a maximum SNR of ~80-100 is 1,000,000 hours for both the Earth and Jupiter. Since an Earth transit takes 13 hours and a Jupiter transit lasts approximately 30 hours, this would require 77,000 and 33,800 transits respectively. Given the orbital periods of Earth and Jupiter then the total observation time to achieve this level of precision for the Earth would be 77,000 years and for Jupiter, 402,000 years. This is clearly not feasible!

Since it is the photon noise from the host star that swamps the planetary signal, we looked to see how detectable the Earth and Jupiter might be in secondary eclipse should they be orbiting a smaller, dimmer M-dwarf. Using a typical M-dwarf spectrum from the Kurucz catalogue, with $R = 0.14 R_{Sun}$ and $M = 0.123 M_{Sun}$ and $L = 0018 L_{Sun}$ we moved Jupiter and the Earth to orbital distances from the host star that would lead to the same radiative equilibrium temperature (0.22 and 0.04 AU respectively) and compared the planetary flux to the host star spectrum and to the expected noise levels for, again, 1, 1000, and 1,000,000 hours transit integration (also Fig. 5). As can be seen, there is some improvement in the detectability of such planets, but it is not dramatic. Similarly, the signal-to-noise (SNR) ratio can be improved if the resolving power is reduced. Fig. 5 also shows the simulation with the Sun as the host star and the FWHM of the instrument function increased from 0.025 μm to Δλ = 1μm. As can be seen the SNR is again improved, but once more the improvement is not especially dramatic and any improvement comes at the expense of reduced spectral discrimination, which would make retrieving the conditions in these atmospheres much more difficult.

Since the integration time required to observe Earth or Jupiter in secondary eclipse is so long for the Sun or M-Dwarfs, there is little reason to present the results of our test retrievals as these measurements are unfeasible. However, for interest test retrievals were performed for the case of 1,000,000 hours integration (giving a peak signal to noise ratio of ~50) and we found that our retrieved atmospheric conditions in all the test cases agreed within retrieval error with the 'true' atmospheric state for Earth, validating our retrieval model and our very simple atmospheric model. Fig. 6 shows the simulated secondary eclipse spectra (about the Sun) for both Earth and Jupiter after 1,000,000 hours integration together with our best fits to them. For Jupiter, the model was complicated by the fact that the SNR of the synthetic spectrum exceeds 10 at wavelengths greater than 8-10 μm and also reaches values of approximately 20 in the 5-micron window, which is prominent as our model atmosphere was tuned to match the brightest 5-micron NIMS real-time spectrum. This feature is well known in Jupiter's spectrum and is where, in the absence of clouds, the atmosphere has low opacity between the phosphine and methane absorption bands allowing radiances from the 5-8 bar pressure level to reach space. Superimposed on this spectrum are features of ammonia and water vapour, whose abundances can thus be determined. The presence of this 5-µm feature means that the observed spectrum probes abundances at several very different levels in Jupiter's atmosphere from deep below the clouds to high in the stratosphere. Using our initial retrieval assumption that the mixing ratios of different gases such as ammonia were constant with height we found that we were unable to fit the 'measured' spectrum very well since in fact very different deep tropospheric and upper tropospheric ammonia abundances are required. Hence, we had to impose extra constraints on some of the a priori gas abundance profiles to properly fit the measured synthetic spectrum. For water, our a priori profile was limited not to exceed the saturated vapour pressure, while for ammonia and phosphine, the abundance was forced to decrease at altitudes above about the 1 bar level at a fraction (0.3) of the pressure scale height. For the real case of Jupiter we know that this happens because of a mixture of condensation and photolysis. If we were to observe a Jupiter-like exoplanet we would naturally be led to the same conclusion of requiring the stratospheric abundance to be much less in order to gain an acceptable fit, and could thus demonstrate active photochemistry.

## 6. Primary Transit Simulations

Having established that secondary eclipses would be very difficult to use for determining conditions in the atmospheres of Earth and Jupiter, we then looked to see how detectable such atmospheres might be for primary transits, again for a 10-m diameter telescope at a distance of 10 light-years with the same instrumental properties as before. We again assumed: 1) a simple model of the temperature profile with a stratospheric temperature determined by the radiative equilibrium temperature, a tropopause at 0.1 bar and adiabatically temperature at depth; 2) that gases were uniformly mixed; and 3) assumed a simple cloud model with a base height, opacity and scale height consistent with observations. Fig. 7 compares the $A_P/A_S$ spectra of both Jupiter and Earth with the estimated noise spectrum for 1, 1,000, and 1,000,000 hours transit integration. At first sight the SNR looks very good (left hand panel of Fig. 7), but it soon becomes obvious that while even a 1-hour transit observation would easily detect the presence of an Earth, the atmospheric absorption features that need to be detected in order to determine the atmospheric conditions are much smaller. Fig. 7 also compares the atmospheric signal component of $A_P/A_S$ (i.e. $A_P/A_S - \min(A_P/A_S)$) with the noise spectrum and we can see that the SNR of detectable atmospheric features is considerably worse. However, we can also see that things look much more promising than the secondary eclipse case, especially at shorter wavelengths.

We computed the transit integration times needed to obtain a peak atmospheric absorption signal SNR of ~100. For Earth we found that 1000 hours were required and given that an Earth transit lasts ~13 hours, this would require observing 77 transits, leading to a total experiment time of 77 years, which is again not feasible. For Jupiter, however, we find that the observation of a single transit, lasting ~30 hours, could determine the atmospheric absorption signal to a peak SNR of over 400. Such an observation would be achievable, although a telescope would have to be pointed at the Solar system for the precise 30 hours of Jupiter's transit during its 11.9 year orbit about the Sun to observe this and it would be difficult to perform stable observations over such a long period. We will return to this case later.

The case for observing primary transits of solar-system like planets about M-Dwarfs is clear from Fig. 8, where we have shown the $A_P/A_S$ spectra of both Jupiter and Earth with the estimated noise spectrum for 1, 1,000, and 1,000,000 hours transit integration orbiting a typical M-dwarf star. Since an M-dwarf has approximately $1/10^{th}$ the radius of the Sun, the eclipse caused by the passage of a solar-system like planet is much larger. In fact, for a planet as large at Jupiter, whose radius is also approaching $1/10^{th}$ the radius of the Sun a very large SNR is predicted for even a 1 hour transit, although the likelihood of partial transits will be much higher. About our M-dwarf, to achieve the same radiative equilibrium temperature as about the Sun, Earth would have to orbit at a distance of ~0.04 AU and would have a transit time of ~1 hours, while Jupiter would have to orbit at a distance of 0.22 AU and would have a transit time (time for centre of planet to travel across the widest part of the star's disc) of ~2.4 hours. Hence, again the transit spectrum of Jupiter could conceivably be recorded with high precision with a single transit, and with an orbit of ~100 days, there would not be so long to wait for the next one. However, the chances of a complete transit (i.e. where the disc of the planet passes entirely in front of the star) are rather small owing to the similarity in size between Jupiter and an M-dwarf. For the Earth, the orbital period would be ~9 days, and a single transit, lasting about 1 hour, would give a peak SNR of ~14. To achieve a peak SNR of ~45 would require 10 hours of integration, or equivalently 10 transits, which would demand an elapsed observation time of only 90 days, which is eminently feasible.

**6.1 Earth M-dwarf transit**

Figure 9 shows the simulated Earth transit spectrum of after co-adding ten M-dwarf transits and our fits to this spectrum (which were all so similar that they are indistinguishable in the figure) using our different initial assumptions of planet radius, stellar radius and planetary mass. The absorption features of a number of gases are visible as peaks in this signal and Figure 10 shows our fitted temperature profile and mean gas abundances compared with the model profiles used to generate the synthetic observation spectrum. The gas abundance profiles plotted here are limited to those where the retrieval finds that the fitted abundance is significantly different from the *a priori* assumptions and the improvement factor (defined as 1 − (retrieval error)/(*a priori* error)) is shown. We should note here that although we retrieve a single abundance at all altitudes, what we are actually retrieving is the mean abundance at the altitude where we have most sensitivity, which varies for the different gases, depending on the position of their spectral features in the observed spectrum. This can be seen in Fig. 11, where we have computed the functional deriviatives (or Jacobians) of the temperature and abundance profiles for our best fit temperature profile and mean abundances. The Jacobians represent the rate of change of signal with respect to a change of abundance, i.e. $dS_j/dx_i$, where $S_j$ is the signal at the j$^{th}$ wavelength and $x_i$ is the abundance of $x$ at the i$^{th}$ vertical level in the atmosphere. For water, the sensitivity was highest from $10^{-1}$ to $10^{-2}$ atm near 6 μm, while for $CO_2$ we are sensitive from $10^{-2}$ to $10^{-4}$ atm at 4.3 and 15 μm. The Jacobians show some sensitivity to temperature (through its indirect effect on the scale height), but not at a level significant enough to move the retrieved temperature profile away from the *a priori*. We can see that the mean gaseous abundances are moderately close to the true values at the vertical level of maximum sensitivity, with the exceptions of $CO_2$, which seems slightly overestimated, and $H_2O$, whose retrieved abundance seems a little high. The Jacobian for $CO_2$ peaks higher in the atmosphere than all other gases and it is likely that an excess of retrieved $CO_2$ is a result of the actual temperature inversion in Earth's atmosphere not being detected. As for $H_2O$, most of the water in Earth's atmosphere is in the lower troposphere, where its absorption features can be masked with cloud. Table 1 compares the 'true' and modelled radius and cloud properties, where we see a reasonably good correspondence between the true and fitted values, although it seems hard to completely disentangle planetary radius from cloud opacity and we find that we consistently overestimate the cloud optical depth. It is thus possible that the $H_2O$ abundance is also being slightly overesitimated to give sufficient absorption above the clouds to match the synthetic 'observed' water feature. The abundance of $O_3$ again appears very retrievable and has a clear feature at 9.6 μm, but the weak $O_2$ feature at 0.76 μm appears undetectable. This is caused by the low resolution of this simulation and also by obscuration due to the uniformly distributed clouds in our simple atmospheric model, which mask out the lower atmosphere (where most of the $O_2$ column abundance resides) in the very long limb pathlengths that light traverses in the primary transit geometry.

**6.2 Jupiter Solar Transit**

Figures 12 and 13 compare the 'true' and modelled spectra and abundance profiles, respectively for a single synthetic Jupiter transit observation of the Sun, while Table 2 compares the 'true' and modelled radius and cloud properties. The gas abundance profiles plotted here are again limited to those where the retrieval finds that the fitted abundance is significantly different from the *a priori* assumptions and the improvement factor is shown. The Jacobians for the temperature profile and abundances are shown in Fig. 14. In this case, while the SNR spectrum again peaks at short wavelengths, there is significant detectability around the $CH_4$ absorption features in the near-infrared and near 8 μm, and also to the absorption features of $C_2H_2$, $C_2H_6$ and $CH_4$ absorption features at 12-14 μm. This sensitivity means that the stratospheric abundance of these gases can be determined, but also, surprisingly, that the increasing temperature in the stratosphere can be determined indirectly through the effects of temperature on the atmospheric scale height, $H$, in Jupiter's atmosphere through the relation $H = RT/Mg$, where $T$ is the temperature, $M$ is the mean molecular weight, $R$ is the Universal Gas Constant, and $g$ is the acceleration due to gravity. The effect is more pronounced for Jupiter's atmosphere than Earth's since the atmospheric scale height is larger (~26 km at the equator, compared with ~9 km for Earth) and Jupiter's atmospheric absorption more vertically extended, with significant absorption up to the 150 – 200 km above the 1 bar level, compared with 30 – 40 km for the Earth. Test retrievals where we fixed the temperature profile to the a priori yielded very poor fits to the synthetic observed spectrum, especially at wavelengths longer than 5 μm. We did wonder, though, if this retrieval might be degenerate in that temperature and hydrocarbon abundances could be compensating for each other, leading to non-unique results. Hence, the retrievals were repeated with 0.1 × the initial a priori gas abundances and then 10 × the initial a priori gas abundances. We found that the retrieved temperature profile and gas abundances all agreed with the nominal case to within the retrieved error. It would thus appear that the stratospheric temperature profile and some gas abundances are clearly distinguishable in this geometry. Thus, while we do not normally expect to be able to infer temperature information from primary transits, it appears that in some cases such as this we actually can. We should note that in the primary transit geometry we are not sensitive to the troposphere since for limb paths the atmosphere 'blacks out' for levels at pressures below the tropopause, especially since we have an extended haze in the middle of the troposphere. Hence, we are insensitive the the abundances of gases such as $NH_3$ and $PH_3$ no matter how long we integrate.

## 7. Directly Observed Spectra

These calculations have shown the difficulty in using transit spectra to measure the atmospheric conditions in cold planets like Jupiter or small terrestrial planets such as the Earth. The thrust behind the space-age exploration of space has, fundamentally, been the search to find life, or the conditions that might give rise to life, elsewhere in our universe. While Hot Jupiters and Hot Neptunes provide fascinating examples of how planets may form which are very different from those seen in our solar system, what really excites our collective imagination is the hope of eventually finding an Earth-like planet elsewhere in our galaxy. We have seen that transit observations are capable of detecting the presence of an Earth-like planet, but are less capable of measuring the spectrum with sufficient precision to determine the atmospheric temperature and composition. Hence, it will be difficult to use such observations to search for the spectral signatures that might indicate the presence of life, unless an Earth-like planet can be viewed in primary transit about a smaller M-dwarf star (although then we would need to consider the harmful effects of the high stellar activity of such stars on the evolution of life). For both primary transits and secondary eclipses it is the photon noise of the host star itself that swamps the signal we want to measure. If we could remove that light, then the task of characterising the atmospheres of these planets would be made much easier. One way of circumventing this problem is direct imaging. Some exoplanets have already been directly imaged and their spectra modelled (e.g. Bonnefoy et al. 2013; Currie et al. 2013; Konopacky et al. 2013; Lee, Heng and Irwin, 2013) although such exoplanets to date have more in common with brown dwarfs than they do with solar system terrestrial planet analogues, for which direct imaging is much more challenging.

Assuming complete cancellation of sunlight we calculated the integration time necessary to measure the flux spectra of Earth and Jupiter in direct imaging mode to a SNR > 100 using our photon noise model and found that just 30 minutes is thoeretically sufficient. In practice, however, many detection systems do not meet the photon noise limit and to achieve perfect nulling is very difficult. However, assuming the best currently available estimates of the noise perfomance of present-day detector technologies and assuming complete cancellation of the Sun's spectrum, Fig. 15 compares the flux spectra of Earth and Jupiter (in units of power, or electron/s, per pixel covering 0.025 μm of the spectral range), with the noise spectrum that could be achieved with just 1 hour's integration and shows that the spectra of both Earth and Jupiter could be well determined.

One way to directly image the Earth and Jupiter from 10 light-years away would be to use nulling interferometry, such as with the proposed ESA/Darwin (Léger et al., 2007; Cockell et al., 2009a,b) or NASA/TPF-Interferometer (Lawson et al., 2009) missions. Such mission proposals use very long baseline interferometry to interferometrically cancel out the light from the star, leaving just the emission from orbiting planets. For reference the maximum elongation of Jupiter from the Sun for an observer at 10 light-years is 1.6 arcsec, while for the Earth it is 0.3 arcsec. Figure 15 also shows the photon noise of the Sun if it could be nulled by a factor of $10^5$ as was proposed for Darwin and we see that while observations are more difficult in the near-infrared, atmospheric absorption features could be detected for both planets by mid-IR observations with sufficient integration. We should stress here that we have assumed 10 m diameter telescopes, whereas the Darwin proposal assumed smaller 2-3 m telescopes and was thus less optimistic about the possibility of detecting earth-like planets (Cockell et al., 2009a,b).

The other main proposed method of direct imaging solar-system-like planets is to physically shade out the light from the central star, either with a space-borne occulter working in conjuction with a separate space telescope, such as the New Worlds Mission (Cash et al., 2009), or with coronography techniques such as the NASA/TPF-Coronograph (Levine et al., 2009) or more recently ATLAST (Postman et al., 2012). If such a mission were to go ahead it is likely that it would operate at visible wavelengths and current state-of-the-art coronograph designs can null the star light by a factor $10^{-10}$ within their optimal range of working angles. At this level of performance, the detection of Earth-like absorption features becomes potentially feasible and proposed space telescopes such as ATLAST are developing these concepts with 10-m class telescopes that could potentially be used to search for Earth-like analogues. Fig. 16 shows the Earth spectrum at this range, which shows absorption features of water, $CO_2$, $O_3$, and molecular oxygen ($O_2$) at 0.76 μm. It can be seen that at our nominal resolution of 0.025 μm, the $O_2$ feature is difficult to distinguish. Hence, we also ran our calculations at ten times the spectral resolution (0.0025 μm) to give spectra which are compared with the original lower resolution spectra in Fig.16. At this resolution the $O_2$ feature is clearly detectable, assuming we could achieve sufficient nulling. Figure 16 also shows the photon noise of the Sun assuming $10^5$ nulling and we can see that the noise is considerably greater than the signal and likely instrument noise and that this problem naturally amplifies at the higher spectral resolution needed to discriminate the $O_2$ absorption. However, if a nulling factor of $10^{10}$ could indeed be achieved it can be seen that we could easily detect the $O_2$ absorption feature at this wavelength with a 10m telescope. It goes without saying that it would be profoundly interesting if we could detect such an $O_2$ signal in the directly imaged spectrum of an exoplanet.

Although direct observations of solar-system-like planets are likely to be challenging in practice, and the Darwin and TPF missions studies were abandoned on the grounds that they are not technically and financially achievable, technology is improving rapidly. Concept studies such as ATLAST suggest that, perhaps quite soon, the case for such a mission could be convincingly made. This study supports the view that such a mission provides the best method for characterising the atmospheric properties of an Earth-like planet orbiting a Sun-like star.

**8. Conclusions**

We have shown in this paper, that were it possible to observe the primary transit and secondary eclipse spectra of planets such as the Earth and Jupiter with sufficient precision, that the bulk properties of their atmospheres could be reliably retrieved with currently existing retrieval techniques, assuming limited prior knowledge. We have also shown that for Earth, the spectral absorption features of $O_3$ are retrievable in the primary transit geometry, should such a planet be orbiting an M-dwarf, from which we might infer evidence of life, although we would have no way of determining atmospheric temperature. However, we have also shown that, with the notable exception of Jupiter in the case of primary transit, transit spectroscopy is not an optimal method for measuring the atmospheric properties of solar-system-like planets orbiting sun-like stars since the photon noise of the host star completely swamps the signature spectra of the planets. For the case of the primary transit of Jupiter in front of the Sun we find that both the temperature and hydrocarbon abundances in Jupiter's stratosphere could be determined in a single transit. Putting this result to one side, though, we conclude that to search for the spectral characteristics of the atmospheres of Earth-like planet atmospheres orbiting Sun-like stars we must attempt to remove the light from the host star, through such techniques as coronography or nulling-interferometry. The technical feasibility of such observations is still in development, but this study supports the view that such missions and approaches are very much worth pursuing and could potentially reap rich rewards in the study of planetary science and in the search for extraterrestrial life in our galaxy.

**Acknowledgements**

We are grateful to the United Kingdom Science and Technology Facilities Council for funding this research. Leigh Fletcher was supported by a Royal Society Research Fellowship at the University of Oxford.

**Appendix A**

Supplementary data, including reference noiseless primary transit and secondary eclipse spectra of Earth and Jupiter, associated with this article can be found at: http://users.ox.ac.uk/~atmp0035/exodata/irwin_2014_ref_spectra.txt .

**Tables**

Table 1. Retrieved properties from Earth primary M-dwarf transit simulation. Here Radius is the planetary radius at the ground, Tau is the nadir cloud optical depth (at 1 μm), and H is the altitude at the base of the cloud. The cloud was assumed to be thin with a fractional scale height equal to 0.1. Comparing the *a priori* with the retrieval error gives an indication of the sensitivity that the retrieval has to the parameter in question, since if there is no sensitivity then the retrieval error is the same as the a priori error. Here we can see that we are very sensitive to Radius, not sensitive to the cloud base altitude and slightly sensitive to the cloud optical depth.

| | Nominal | Mass Reduced | Mass Increased | Radius Reduced | Radius Increased | Reduced Star Radius | Increased Star Radius | A priori error | Retrieval Error | True |
|---|---|---|---|---|---|---|---|---|---|---|
| Radius (km) | 6371.0 | 6369.0 | 6689.2 | 6372.3 | 6052.7 | 6054.6 | 6686.8 | 100. | 1.0 | 6378.14 |
| Tau | 0.46 | 0.62 | 0.46 | 0.38 | 0.43 | 0.32 | 0.65 | 0.1 | 0.03 | 0.1 |
| H (km) | 5.1 | 5.3 | 5.1 | 5.1 | 5.0 | 5.0 | 5.4 | 1.0 | 1.0 | 5 |

Table 2. Retrieved properties from Jupiter primary transit simulation. Here Radius is the planetary radius at the 1 bar level, Tau is the nadir cloud optical depth (at 1 μm), and H is the altitude at the base of the cloud. The cloud was assumed to be extended with a fractional scale height equal to 0.5. Comparing the *a priori* with the retrieval error we can see that we are again very sensitive to Radius, and moderately sensitive to the cloud base altitude and cloud optical depth.

|  | Nominal | Mass Reduced | Mass Increased | Radius Reduced | Radius Increased | Reduced Star Radius | Increased Star Radius | A priori error | Retrieval Error | True |
|---|---|---|---|---|---|---|---|---|---|---|
| Radius (km) | 71489.7 | 71486.9 | 71492.4 | 71489.1 | 75060.3 | 67918.2 | 71489.7 | 100. | 0.5 | 71492.0 |
| Tau | 5.49 | 7.80 | 5.92 | 6.60 | 8.02 | 4.35 | 5.49 | 5.0 | 1.5 | 5.2 |
| H (km) | 11.8 | 10.7 | 10.2 | 11.5 | 11.8 | 10.6 | 11.8 | 2.0 | 1.0 | 12.95 |

**Figures**

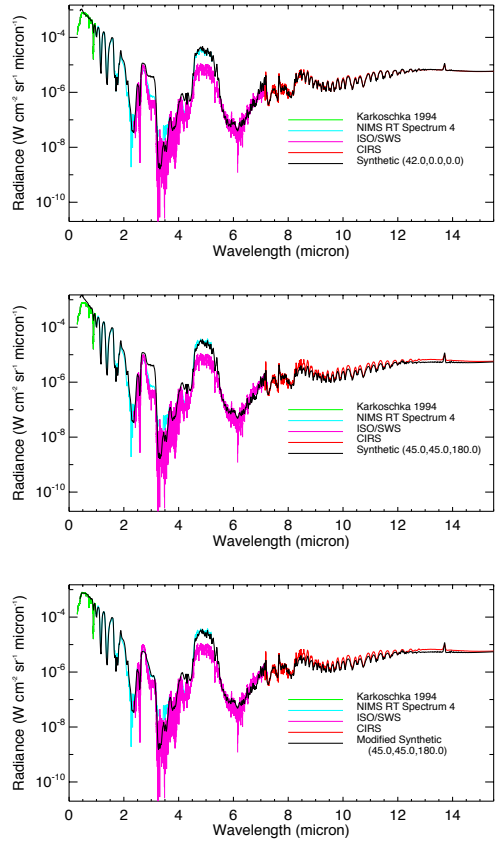

Figure 1. Comparison between measured (NIMS, ISO/SWS, CIRS and ground based (Karkoschka 1994)) and our calculated synthetic Jupiter radiance spectrum for different observing geometries and assumed particle scattering characteristics as described in the text. The top panel compares measured spectra with our calculation at the NIMS observing geometry of solar zenith angle = 42°, viewing zenith angle = 0° and azimuth angle = 0°. The middle panel compares the observations with a more representative disc-averaged case (most consistent with ISO/SWS) of solar zenith angle = 45°, viewing zenith angle = 45° and azimuth angle = 180° . The bottom panel also compares the model and observations in the disc-averaged case, but here the single scattering albedo has been adjusted in the 0.4-0.7 μm and 2.5-3.0 μm regions to obtain a better agreement between the model and measurements.

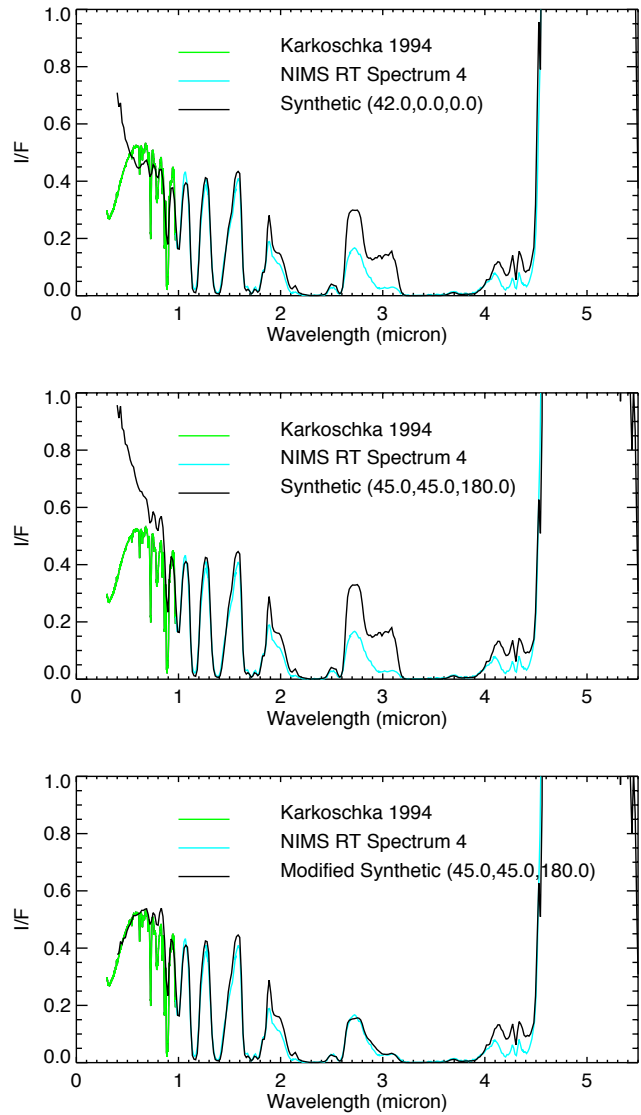

Figure 2. Comparison between measured and synthetic Jupiter reflectivity (I/F) spectra for different observing geometries and assumed particle scattering characteristics. Again, the top panel compares measured spectra with our calculation at the NIMS observing geometry of solar zenith angle = 42°, viewing zenith angle = 0° and azimuth angle = 0°. The middle panel compares the observations with a more represetive disc-averaged case (most consistent with ISO/SWS) of solar zenith angle = 45°, viewing zenith angle = 45° and azimuth angle = 180° . The bottom panel is also for the disc-averaged case, but here the single scattering albedo has been adjusted in the 0.4-0.7 μm and 2.5-3.0 μm regions to obtain a better agreement.

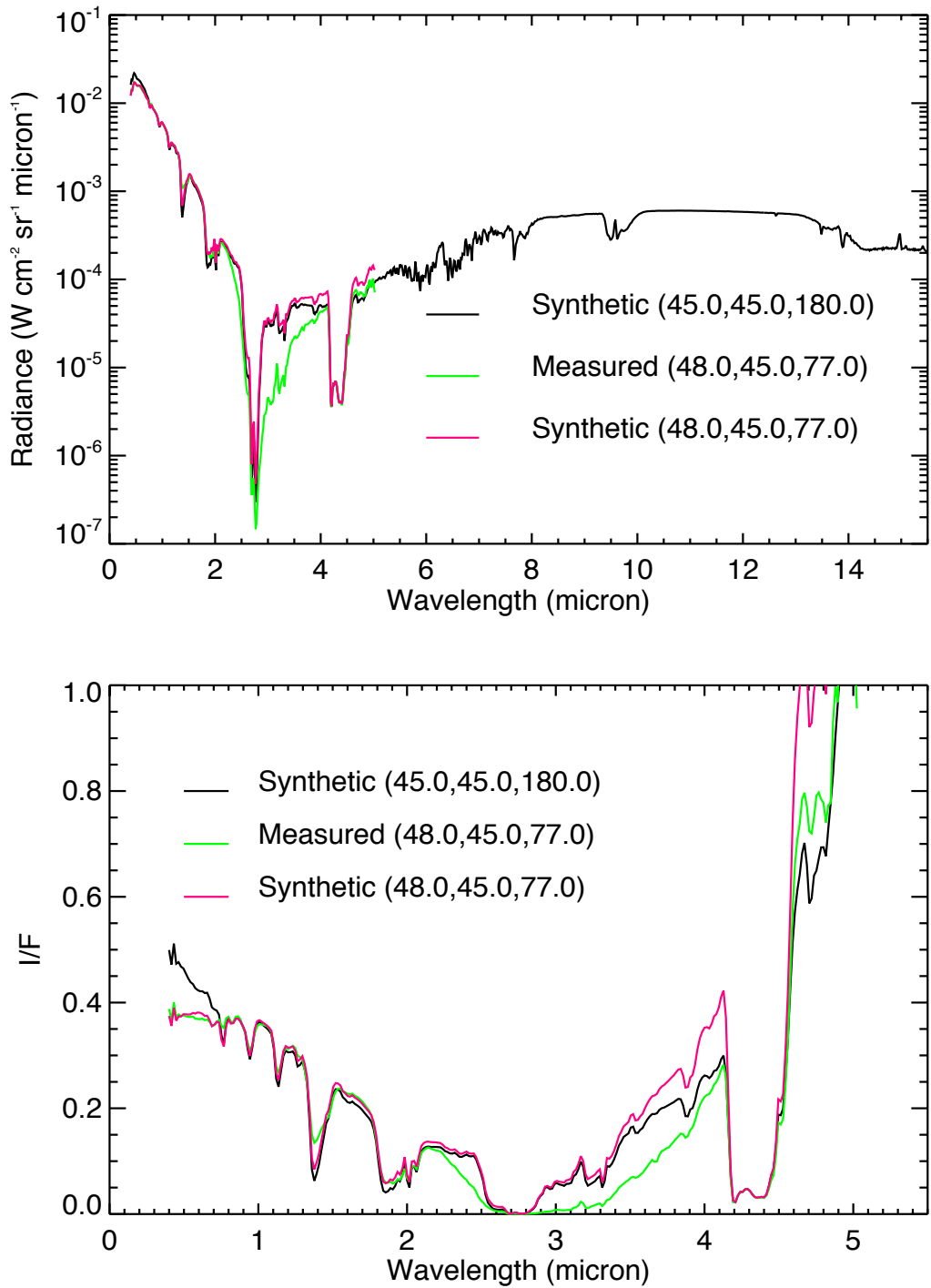

Figure 3. Comparison between measured and synthetic Earth near-infrared radiance and reflectivity (I/F) spectra for the disc-averaged observing geometry of solar zenith angle=45°, viewing zenith angle=45°, azimuth angle=180°, and also the mean geometry of the Rosetta/VIRTIS disc observations.

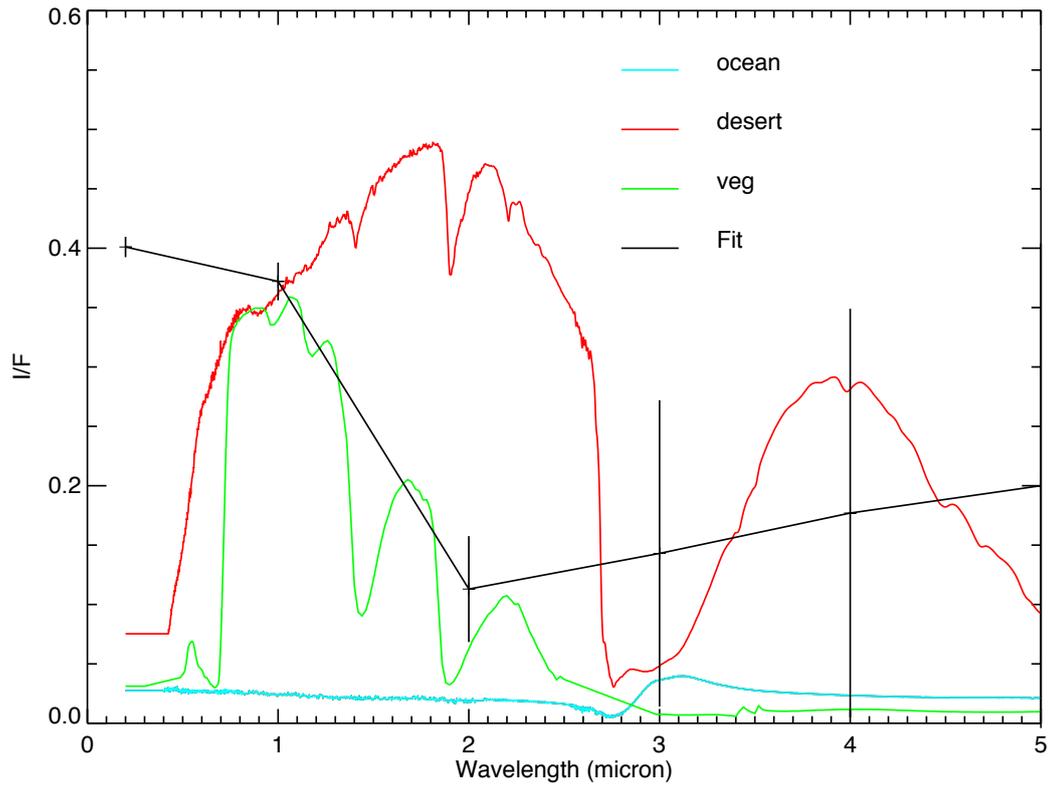

Figure 4. Retrieved near-infrared mean surface albedo spectrum of the Earth compared with typical albedo spectra of representative surfaces extracted from the JPL ASTER Spectral Library (http://speclib.jpl.nasa.gov/search-1). The a priori albedo was set to 0.2±0.2 at all near-IR wavelengths. It can be seen that the solution has not moved very far from the a priori at wavelengths longer than 2 μm.

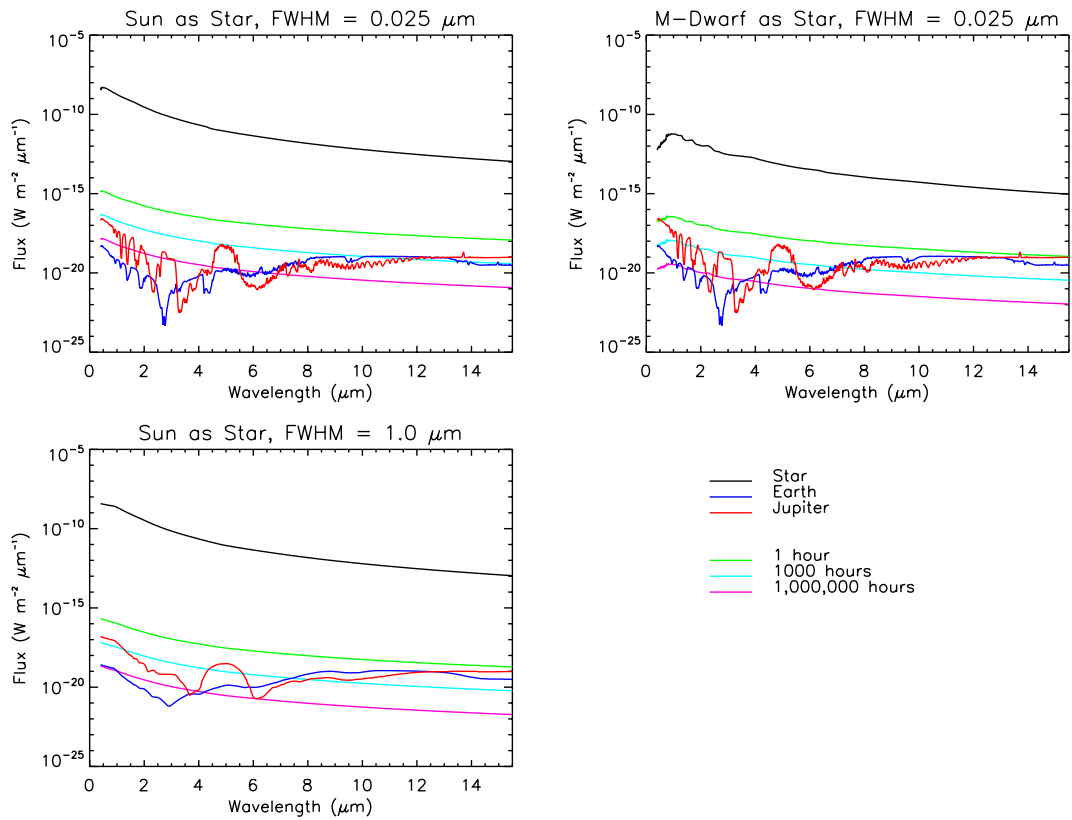

Figure 5. Disc-integrated flux spectra of Earth and Jupiter compared with that of the Sun and also an M-dwarf. The noise levels for different periods of integrations are also shown. Panel A (top left) shows the case with the Sun as the host star and a spectral resolution of 0.025 μm. Panel B (top right) shows the same calculation for an M-dwarf, where the planets have been moved closer to the star to achieve the same radiative equilibrium temperature. Panel C (bottom left) shows the calculations with the Sun as the host star again, but with the spectral resolution degraded to 1.0 μm.

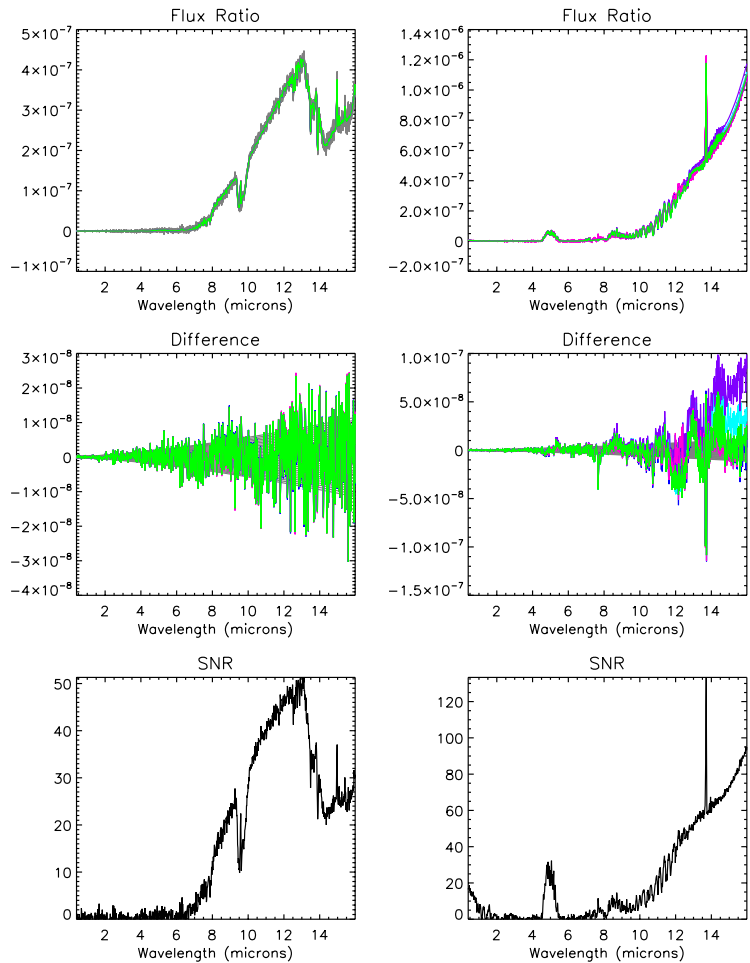

Figure 6 Top row shows a simulated secondary eclipse spectrum of the Earth (left) and Jupiter (right) after $10^6$ hours integration assuming our standard model atmospheres together with the error limits (grey region) and our best fit spectra from our different test cases (coloured or non-solid lines, colours or linestyles are indicated in Fig. 10). Middle row shows the difference between the modelled and synthetic measured spectra (coloured or non-solid lines, corresponding to the different perturbed cases described in the text) together with the estimated noise values (grey region). Bottom row shows the SNR of the synthetic measured spectrum.

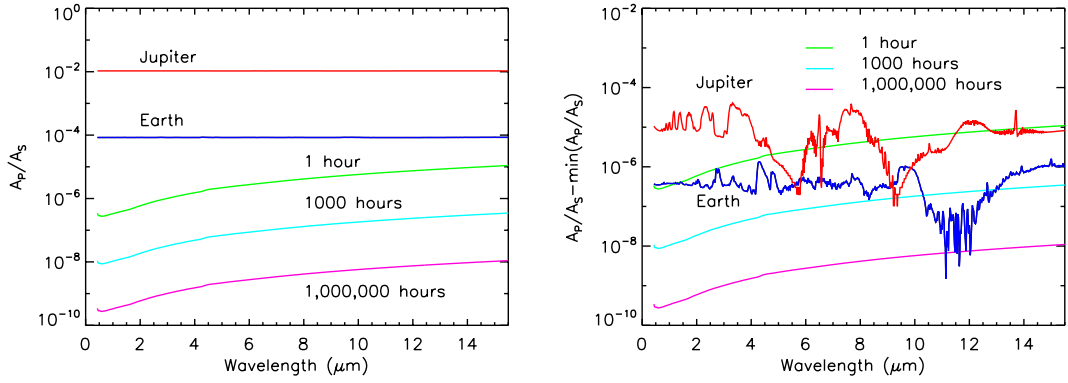

Figure 7. Planet/Star area ratios for the Earth and Jupiter compared with the Sun and also with the precision with which this quantity could be determined from observing primary transits for different integration times. The left hand panel shows that such observations can detect the presence of a solar system target with ease. However, the right hand panel shows that to measure the spectral features in the $A_P/A_S$ ratio that contain information on the atmospheric structure requires considerably longer integration times.

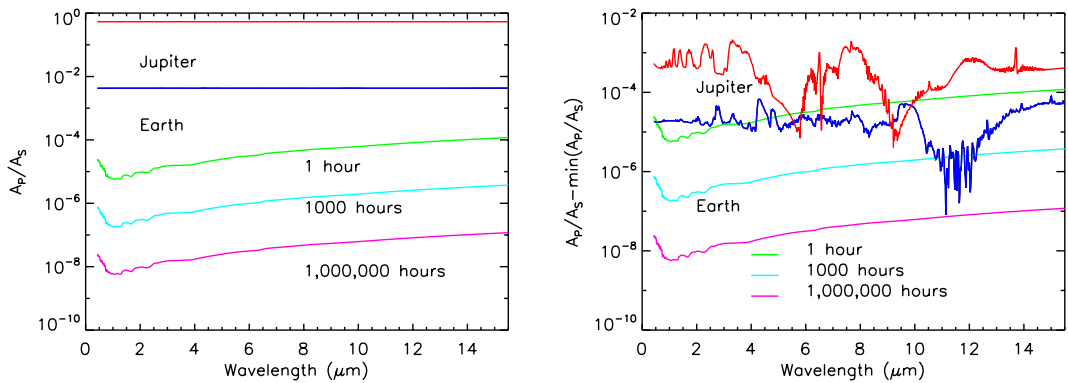

Figure 8. As Figure 7, but for the case of Eath and Jupiter orbiting an M-dwarf. It can be seen that the detectability of atmospheric features is greatly improved over the case when the host star is the Sun.

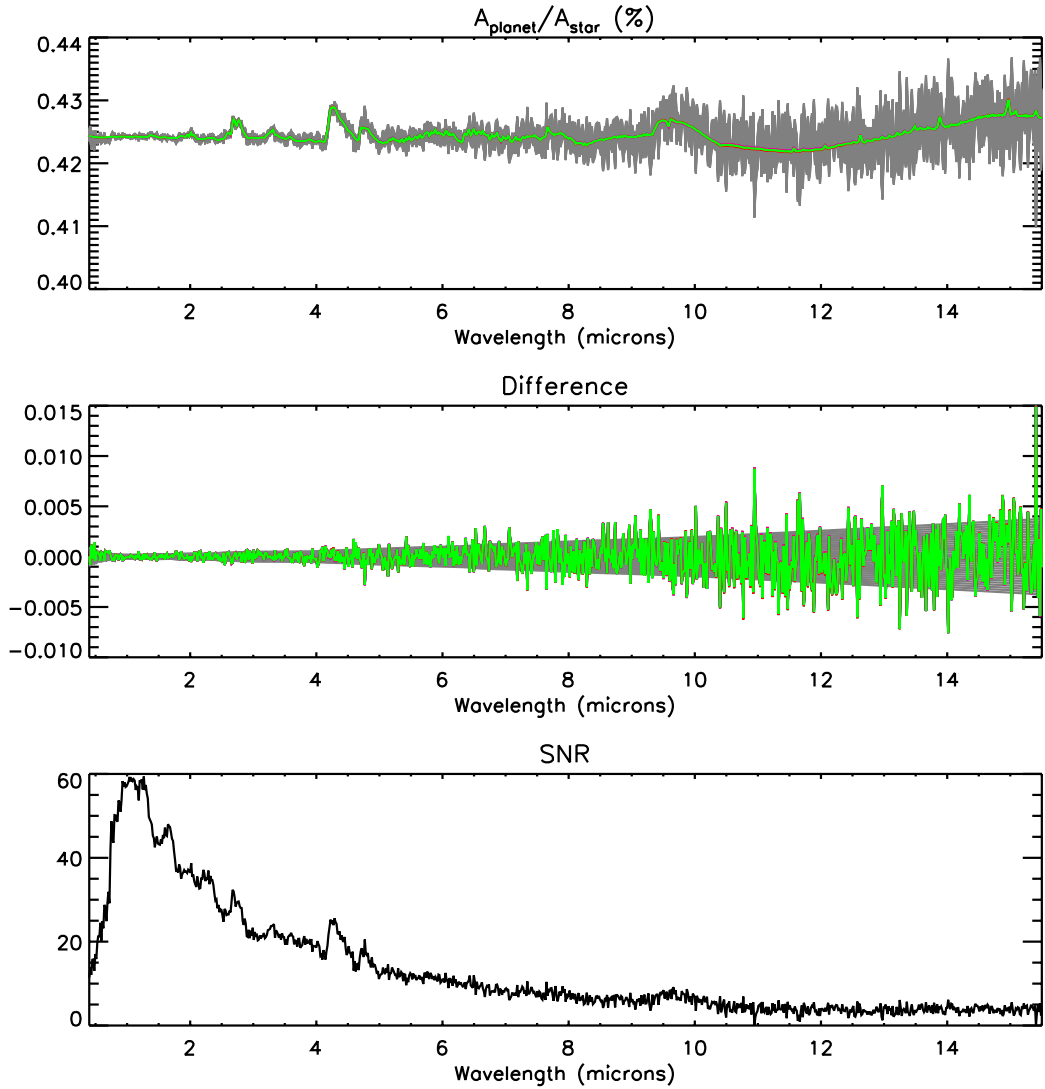

Figure 9. As Fig. 6, but for ten primary transits of the Earth about an M Dwarf (10 hours transit integration time), with our seven different initial assumptions of: 1) nominal; 2) reduced planetary mass; 3) increased planetary mass; 4) reduced planetary a priori radius; 5) increased planetary a priori radius; 6) reduced stellar radius; and 7) increased stellar radius. Note that the fits for the different cases are all so similar that they are indistinguishable. Note also that the SNR is calculated as (spectrum-min(spectrum))/error.

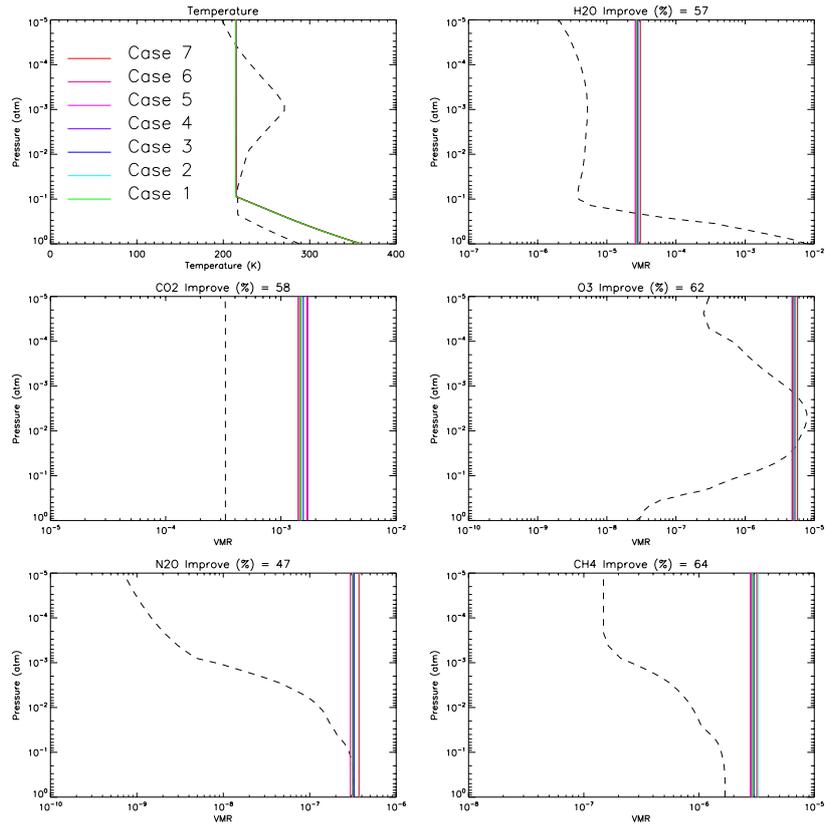

Figure 10 Retrieved profiles from observing ten transits of the Earth about an M-Dwarf. The different retrieval cases are: 1) nominal; 2) reduced planetary mass; 3) increased planetary mass; 4) reduced planetary a priori radius; 5) increased planetary a priori radius; 6) reduced stellar radius; and 7) increased stellar radius. The top left panel shows the retrieved temperature profiles for the different cases (colours or linestyles indicated) compared with the true profile (dashed line) and a priori profile (dotted line). In this case, where there is no temperature imformation, the fitted profiles all lie on top of the a priori profile. The other panels show the mean retrieved abundances of the most retrievable gases with their improvement factors (see main text) quoted in the labels. Here the true profiles are again shown as the dashed lines, while the fitted profiles are shown by the coloured or non-solid lines. It can be seen that the retrieved profiles in all seven cases are almost indistinguishable compared with the true variation of these parameters in Earth's atmosphere.

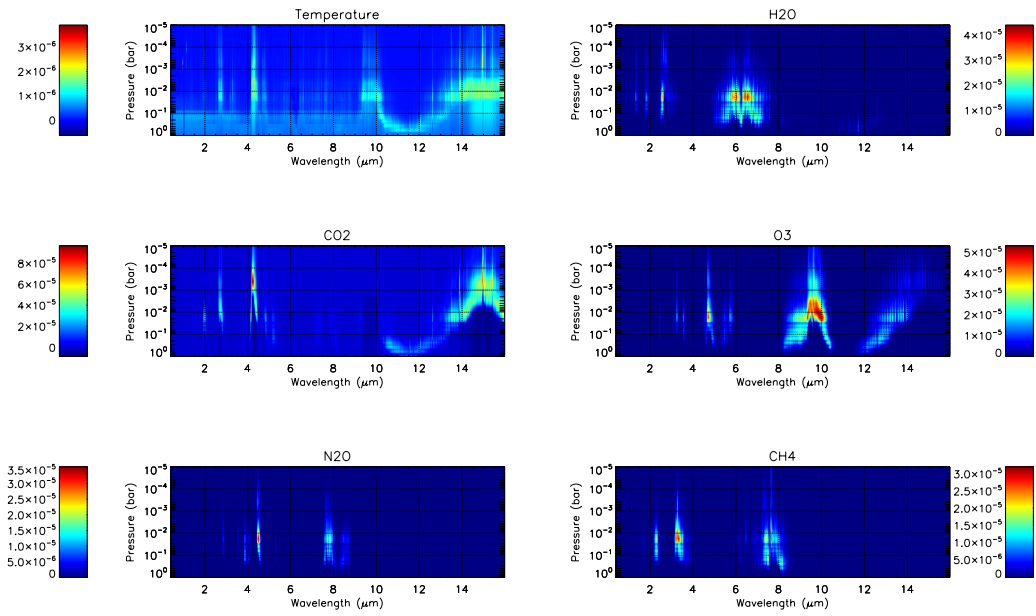

Figure 11. Jacobians (or functional derivatives) for the temperature and abundance profile retrievals for 10 transits of an Earth-like planet about an M-dwarf, showing the position of main absorption features and also the vertical levels at which the Jacobians are most sensitive.

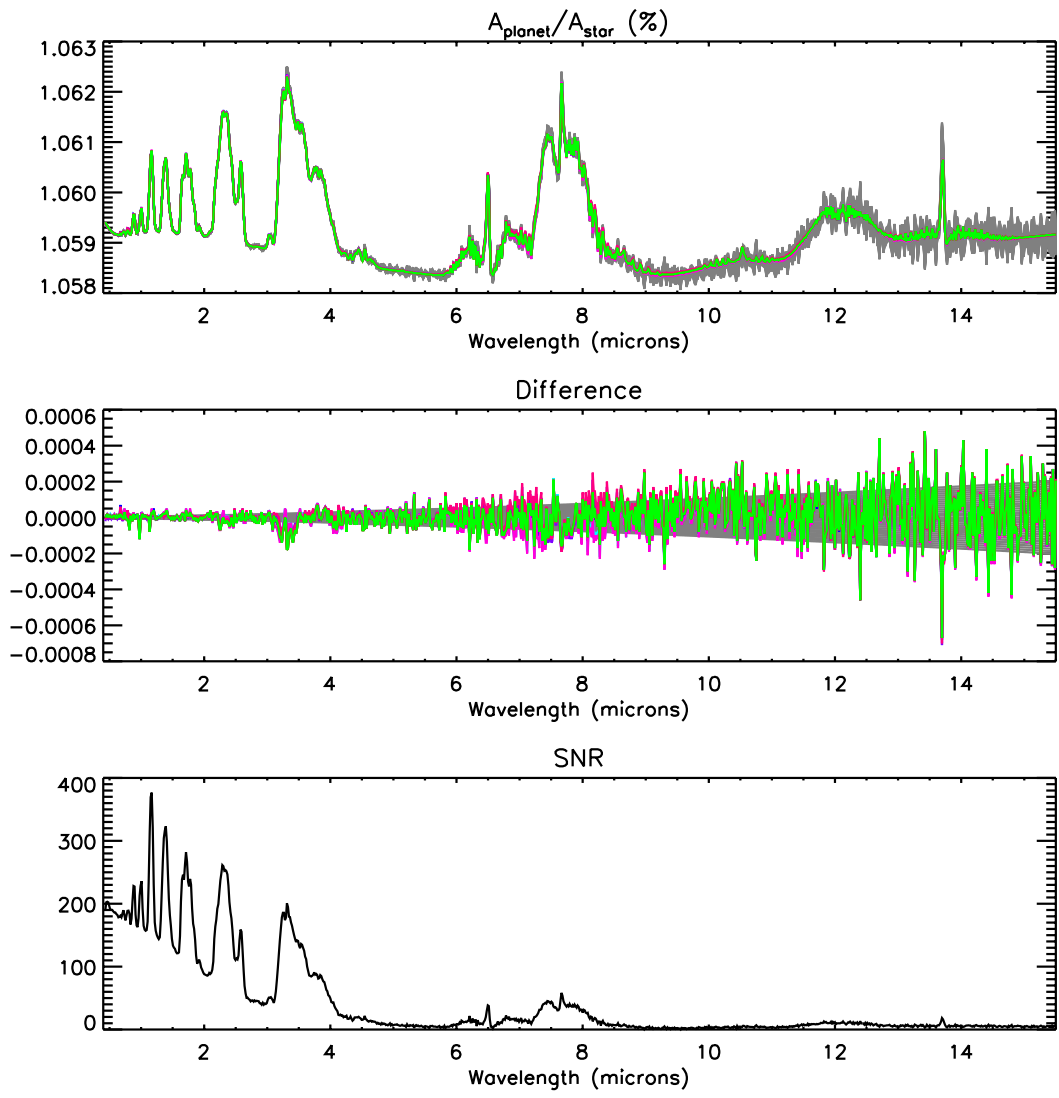

Figure 12. As Fig. 9, but for a single primary transit of Jupiter in front of the Sun.

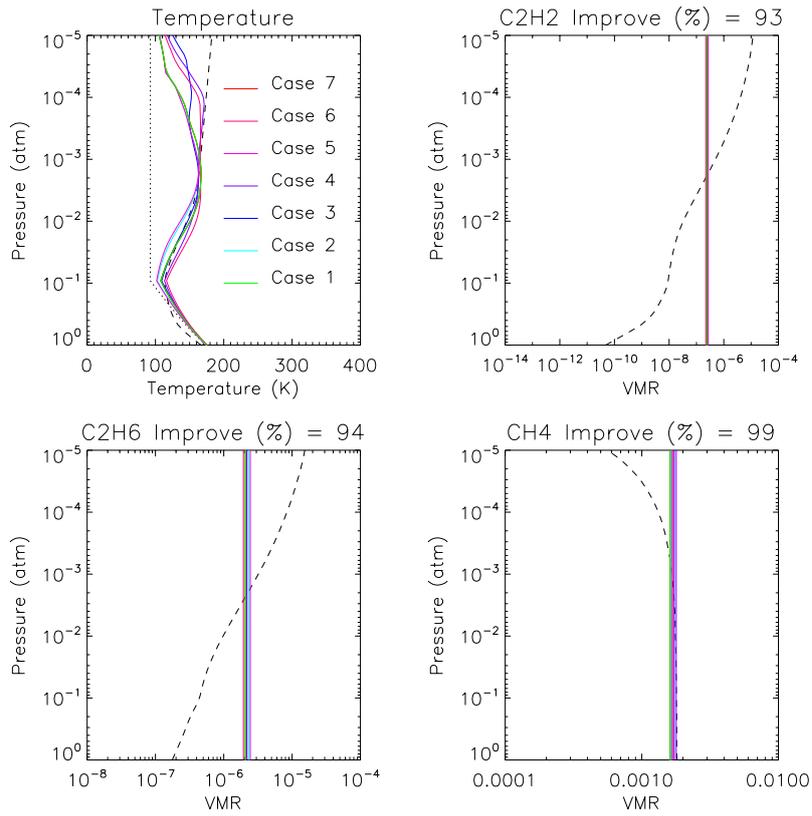

Figure 13. As Fig. 10, but for a single Jupiter primary transit of the Sun. In all plots the dashed lines are the true profiles. For the top left plot, showing the temperature profile, the dotted line is the assumed *a priori* temperature profile.

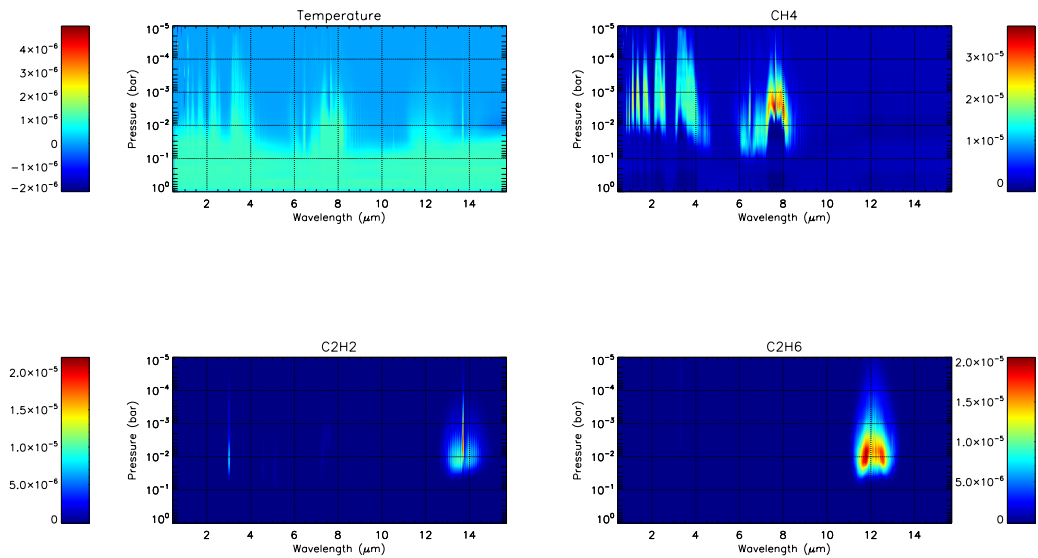

Fig. 14. As Fig.11 but for a single Jupiter transit of the Sun.

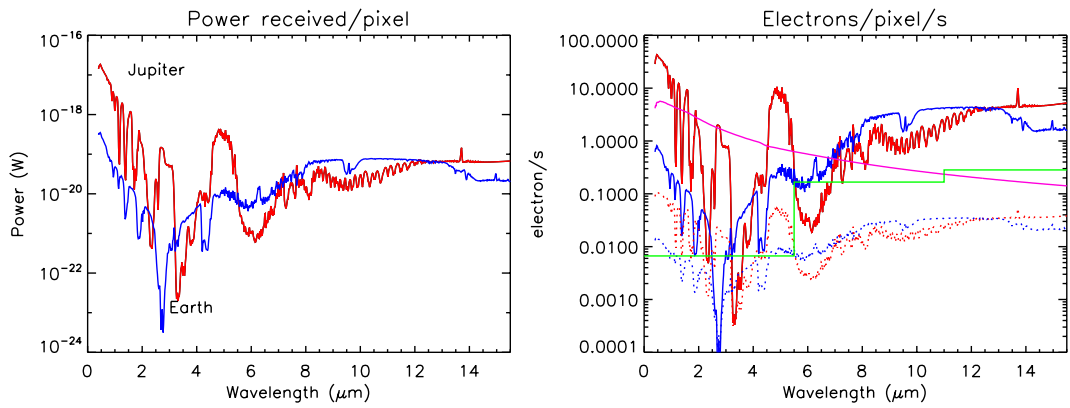

Figure 15. Disc-integrated fluxes of Earth (blue or dark grey) and Jupiter (red or light grey) for direct imaging, converted to power/spectral element (covering 0.025 μm of the spectral range) in the left hand panel for a 10m telescope at 10 light years with τ=0.5, Q=0.7 and Δλ=0.025 μm, integrating for 1 hour. The right hand panel expresses these spectra in terms of electrons/pixel/s together with the noise performance of the best currently available detectors (green or dashed line) and also the photon limited noise values (dotted lines), both for a 1-hour integration. The pink (or solid black) line shows the photon noise from the Sun that arises from incomplete nulling of the Sun's light to a factor of $10^5$.

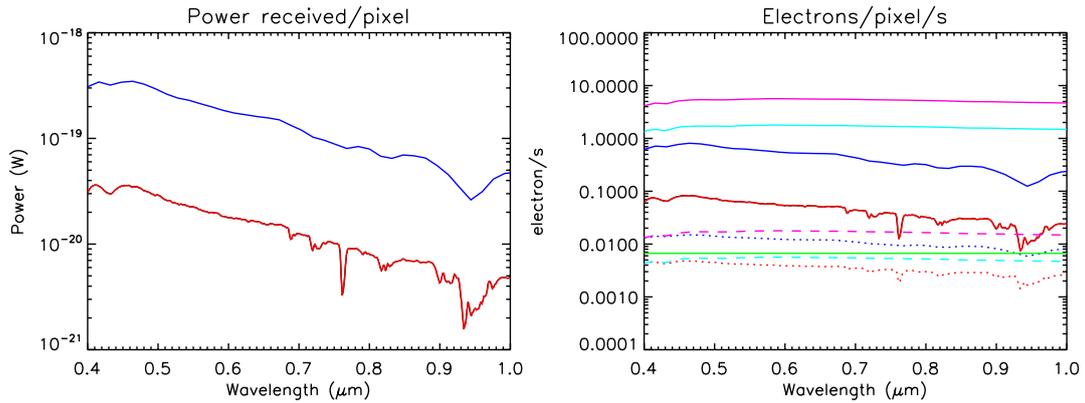

Figure 16. Disc-integrated fluxes per spectral element of Earth for direct imaging at original resolution of Δλ=0.025 μm (blue or dark grey) and at Δλ=0.0025 μm (red or light grey), both observed for 1 hour. As in Fig. 13, the left hand panel shows the power/spectral element for a 10m telescope at 10 light year with τ=0.5, Q=0.7, while the right hand panel expresses these in terms of electrons/pixel/s together with the noise performance of current detectors (green, or long-dashed line) and also the photon limited noise values (dotted lines), both for a 1-hour integration. The right hand panel also shows the photon noise from the incompletely nulled Sun (nulling = $10^5$) at a resolution of Δλ=0.025 μm (pink, or black) and also Δλ=0.0025 μm (cyan, or dashed). In the colour version the dashed lines of the same colour show the Sun's photon noise assuming a nulling of $10^{10}$. In the black and white version the same spectra are indicated with dash-dot and dash-dot-dot lines. The shallow absorption caused by the Chappuis ozone bands can just be discerned, centred on 0.55 μm, at both resolutions while the $O_2$ line at 0.76 μm is only clearly distinguishable at the Δλ=0.0025 μm resolution.